\documentclass[twocolumn,showpacs,preprintnumbers,amsmath,amssymb]{revtex4}

\usepackage{graphicx} 
\usepackage{dcolumn} 
\usepackage{bm} 

\begin{document}

\preprint{}

\title{Charge Pumping in Carbon Nanotubes}

\author{P. Leek}
\altaffiliation{These authors contributed equally to this work}
\author{M.R. Buitelaar}
\altaffiliation{These authors contributed equally to this work}

\author{V.I. Talyanskii}
\author{C.G. Smith}
\author{D. Anderson}
\author{G. Jones}

\affiliation{Cavendish Laboratory, University of Cambridge,
Madingley Road, Cambridge, CB3 0HE, UK}

\author{J. Wei}
\author{D.H. Cobden}

\affiliation{Department of Physics, University of Washington,
Seattle, Washington}

\date{\today}

\begin{abstract}
We demonstrate charge pumping in semiconducting carbon nanotubes
by a traveling potential wave. From the observation of pumping in
the nanotube insulating state we deduce that transport occurs by
packets of charge being carried along by the wave. By tuning the
potential of a side gate, transport of either electron or hole
packets can be realized. Prospects for the realization of nanotube
based single-electron pumps are discussed.
\end{abstract}


\maketitle


The phenomenon of charge pumping has attracted considerable
interest in the last two decades from both fundamental and applied
points of view
\cite{Thouless,Niu,Geerligs,Kouwenhoven,Talyanskii1,Switkes,Keller,Talyanskii2,Sharma,Ebbecke}.
In pumping, a periodic in time and spatially inhomogeneous
external perturbation yields a dc current. If a fixed number $n$
of electrons is transferred during a cycle then the pumping
current is quantized in units of $ef$, where $e$ is the electron
charge and $f$ is the perturbation frequency. An important aspect
of single-electron pumps is their potential to provide an accurate
frequency-current conversion which could close the measurement
triangle relating frequency, voltage, and current. Previously, a
realization of quantized current $I = nef$ has been achieved in
two different ways: first, using devices comprising charge islands
and controlled by a number of phase-shifted ac signals
\cite{Geerligs,Kouwenhoven,Keller}; and second, using
one-dimensional (1D) channels within a GaAs heterojunction where a
surface acoustic wave (SAW) produces traveling potential wells
which convey packets of electrons along the channel
\cite{Talyanskii1}. In the SAW pumps, transport of charge
resembles the pumping of water by an Archimedean screw. When this
principle is combined with Coulomb blockade it results in the
pumping of a fixed number of electrons $n$ per cycle. For
metrological applications, the delivered current should be in the
range of 1 nA and at present only the SAW single-electron pumps
satisfy this requirement. However, the accuracy of the SAW pumps
must be improved significantly for them to find metrological
applications.

A quantum regime of pumping, in which quantum interference plays a
key role, was first described by Thouless \cite{Thouless,Niu}. In
the Thouless mechanism, a traveling periodic perturbation induces
minigaps in the spectrum of an electronic system, and when the
Fermi level lies in a minigap an integer number of electrons $n$
are transferred during a cycle, resulting in a quantized current
flowing without dissipation. From a fundamental physics
standpoint, this mechanism represents a new macroscopic quantum
phenomenon reminiscent of the quantum Hall effect and of
superconductivity. Possible applications of charge pumping are not
limited to metrology. For example, the ability of the pumps to
control the position of single electrons could be used in various
quantum information processing schemes \cite{DiVincenzo,Barnes}.

\begin{figure}
\includegraphics[width=75mm]{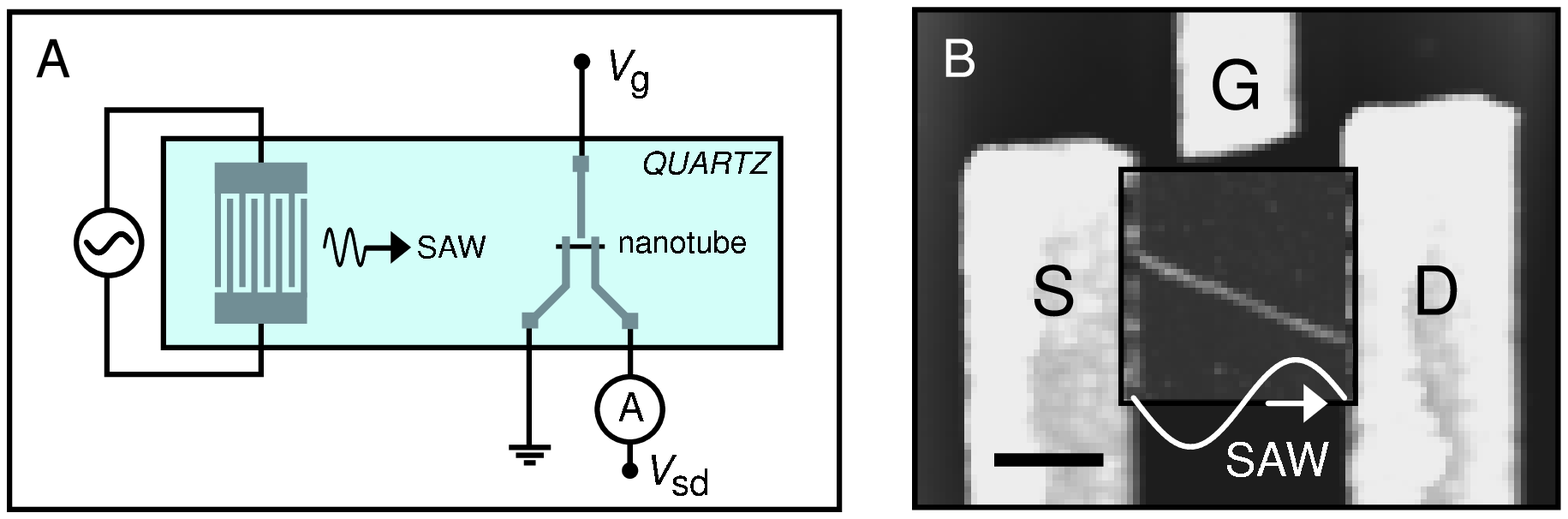}
\caption{\label{Fig1} \textbf{(a)} Schematic of the device showing
the transducer and carbon nanotube. \textbf{(b)}AFM image of the
contacted nanotube. Scale bar is 500 nm. The grayscale height
range is smaller for the central square to highlight the nanotube.
The SAW pumps charge from the source (S) to the drain (D), while
carrier type is controlled by the nearby gate (G).}
\end{figure}

Recently it has been pointed out that carbon nanotubes have
significant advantages over semiconductor and metallic systems in
terms of single-electron pumping \cite{Talyanskii2,Sharma}. The
typical Coulomb charging energies achievable in nanotubes can
exceed 10 meV, significantly larger than in the pumps described in
refs. \cite{Geerligs,Kouwenhoven,Talyanskii1,Keller}. Carbon
nanotubes also offer advantages in the quantum regime of pumping
due to an order of magnitude larger achievable minigaps for a
given spatial period of the perturbation \cite{Talyanskii2}.
Motivated by the potential of nanotubes in this regard, we have
undertaken an experimental search for charge pumping in nanotubes
by the traveling potential of a SAW. Here we demonstrate that a
SAW can indeed pump charge through a carbon nanotube and that the
amplitude and polarity of the current can be controlled by a
nearby gate electrode. In the final part of the paper we discuss
possible applications of the discovered effects.

We employ the arrangement suggested in \cite{Talyanskii1} and
shown in Fig.~\ref{Fig1}(a). A nanotube (grown by chemical vapour
deposition) lies on a polished $36^{\rm{o}}$ Y-cut quartz
substrate. Source and drain contacts (separated by 1 $\mu$m), a
side gate, and a transducer for SAW generation are fabricated by
electron beam lithography. The transducer is 0.7 mm away from the
nanotube and is designed to generate a SAW with a wavelength of 1
$\mu$m corresponding to a SAW frequency of 3.2 GHz. Since quartz
is piezoelectric, the SAW is accompanied by an electrostatic
potential wave which acts on the electrons in the nanotube. All
the data presented here are for a nanotube of diameter \mbox{$2.5
\pm 0.5$\,nm} at \mbox{$T = 5$\,K} \cite{diameter,devices}.

\begin{figure}
\includegraphics[width=75mm]{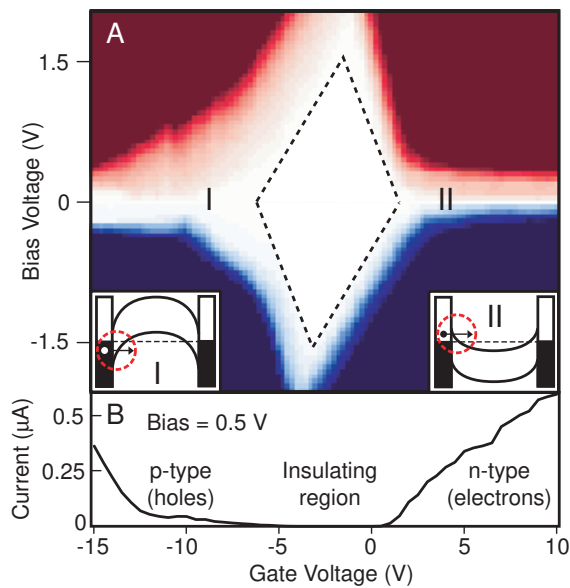}
\caption{\label{Fig2} \textbf{(a)} Colorscale plot of the dc
current versus gate and source-drain bias voltage at \mbox{$T =
5$\,K}. Red indicates positive and blue negative currents. The
superimposed rhomboid, taken at a threshold current of 0.5 nA,
gives an indication of the location of the insulating region.
(Insets) Schematic diagrams showing the difference in Schottky
barrier height presented to electrons and holes. \textbf{(b)} Line
scan taken at \mbox{$V_{sd} = 0.5$\,V}.}
\end{figure}

Fig.~\ref{Fig2}(a) shows a color scale plot of the current as a
function of source-drain bias ($V_{sd}$) and gate voltage ($V_g$)
in the absence of a SAW excitation. A rhomboid-shaped region of
low conductance is observed in the center (-5 V $< V_g <$ 1 V),
indicating that this nanotube is semiconducting. The slightly
higher conductance in the $n$-type regime than in the $p$-type is
consistent with the expected difference in Schottky barrier
heights for electrons and holes at the Ti contacts \cite{Derycke}.
More resolved measurement (not shown) in the $n$ and $p$-type
regions show Coulomb blockade oscillations with a charging energy
of order \mbox{$5-10$\,meV}.

\begin{figure}
\includegraphics[width=75mm]{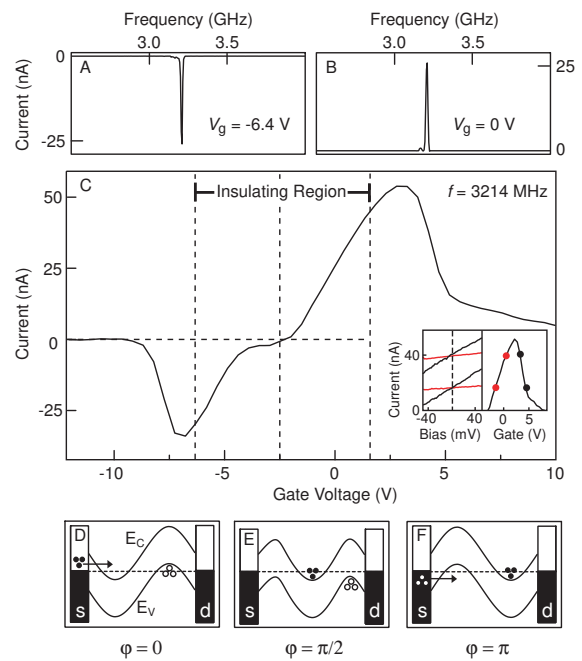}
\caption{\label{Fig3} \textbf{(a,b)} Dependence of the SAW-induced
current on RF frequency for two values of the gate voltage,
showing distinct peaks at the resonant SAW frequency. RF power
applied to the transducer is +20 dBm. \textbf{(c)} Dependence of
the SAW-induced current on gate voltage. Positive current
corresponds to electron flow in the direction of SAW propagation.
(Inset) Dependence of the current on dc bias for four fixed values
of the gate voltage, showing a transition from low to high
conductance at the current peak. RF power is +20 dBm.
\textbf{(d-f)} Schematic band diagrams depicting the proposed
mechanism of charge transport. The conduction ($E_C$) and valence
($E_V$) bands are bent by the SAW allowing injection of electrons
or holes depending on the phase ($\phi$) of the wave
\cite{potential}.}
\end{figure}

\begin{figure}
\includegraphics[width=75mm]{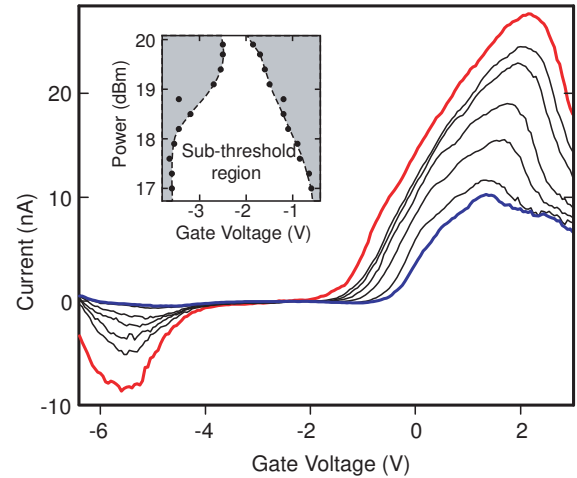}
\caption{\label{Fig4} SAW-induced current versus gate voltage for
fixed values of RF power 17.3 (blue), 17.9, 18.8, 19.1, 19.4,
19.7, and 20 dBm (red). (Inset) Points correspond to
\mbox{$|I_{SAW}|= 0.1$\,nA}. This small threshold current was
arbitrarily chosen to indicate the region in the $P-V_g$ plane in
which a SAW induced current is absent. The dashed lines are a
guide to the eye.}
\end{figure}

Applying RF power $P_{RF}$ to the transducer induces a dc current
$I_{SAW}$ in the nanotube in the absence of applied bias. The
current is generated only within a narrow frequency range,
corresponding to the passband of the transducer, see
Fig.~\ref{Fig3}(a,b). Hence it occurs only when a SAW is present,
and is not a result of rectification of airborne RF fields. Figure
\ref{Fig3}(c) shows the variation of $I_{SAW}$ with $V_g$ at
\mbox{$P_{RF} = 20$\,dBm} (100 mW). At this power level the
current is present across the entire gate region in which the
low-bias dc conductance is absent, see Fig.~\ref{Fig2}(b),
reversing direction in the center and showing a peak located just
outside the insulating region in each direction. At lower power
levels there is a range of $V_g$ in which $I_{SAW}$ is zero, as
can be seen in Fig.~\ref{Fig4}. As the power level decreases this
range widens and the peak current in each direction decreases
steadily.

For the interpretation of the results in Figs.~2-4 it is
convenient to start by observing that the pumped current flows in
the insulating state of the nanotube (-5 V $< V_g <$ 1 V) where
there are no free carriers in the bulk of the nanotube in the
absence of the SAW. It implies that carriers must be injected into
the nanotube from the source contact by the SAW and, once
injected, become trapped in SAW potential minima (or maxima for
holes) and are carried along the nanotube by the wave, see
Fig.~\ref{Fig3}(d-f).

Before discussing the implications of a model of charge transport
in packets, we note that the observation of charge transport in
the insulating state of the nanotube also rules out an alternative
transport mechanism in which a SAW-induced rectified voltage would
drive a current. A rectification mechanism requires the presence
of a region with a nonlinear conductivity. For example, the
nonlinearity of the nanotube contacts could be a natural source
for a rectified dc voltage. The value of such a rectified voltage,
however, would not exceed the SAW amplitude which can be estimated
from the known geometry of the transducers. We obtain
\mbox{$\Phi_{SAW} \sim 0.2$\,V} at 20 dBm of RF power
\cite{Miller}. From Fig.~\ref{Fig2} it follows that when the Fermi
level is close to the centre of the semiconducting gap, a dc
voltage above 1 V is required for a current to flow. This is
significantly larger than any possible rectified voltage which
could therefore not result in the injection of charge into the
nanotube. Moreover, a rectification mechanism could not explain
that in our device the current is not sensitive to a source-drain
bias (see insert of Fig.~\ref{Fig3}(c)). Finally, we note that the
sign of the total rectified voltage is not related to the
direction of the SAW propagation, while in our experiments the
direction of the induced current always followed that of the SAW.

The packet transport model is depicted in Figs.~\ref{Fig3}(d-f).
The figures show schematic band diagrams at the moments when the
electrons and holes are injected for the gate voltage at which the
Fermi level is in the centre of the semiconducting gap in the
absence of the SAW (\mbox{$V_g \sim -2.5$\,V}). Here, we neglect
the relatively small difference in barrier height for the
electrons and holes at the contacts. We also assume that the
injection takes place when the SAW electric field is at its
maximum at the source contact. The SAW bends the conduction and
valence bands of the nanotube so that when the bottom (top) of the
conduction (valence) band is below (above) the Fermi level of the
source contact, electrons (holes) can tunnel into the nanotube
\cite{bandbending}. Thus, the SAW amplitude must exceed a gate
voltage dependent threshold value for the injection of charge to
be possible. With the Fermi level in the centre of the gap, the
thresholds for electron and hole injection are the same, and the
threshold SAW amplitude is related to the semiconducting gap as
$\Phi_{SAW} \textrm{(threshold)} = E_{gap}/2e$. At 20 dBm of RF
power applied to the transducer, the SAW amplitude only slightly
exceeds the threshold (see Fig.~\ref{Fig3}) resulting in an
estimate of \mbox{$E_{gap} \sim 0.4$\,eV}. This is in reasonable
agreement with the expected value of 0.3 eV for semiconducting
nanotubes of diameter \mbox{$\sim 2.5$\,nm} \cite{Wildoer}.

For SAW amplitudes exceeding $\Phi_{SAW} = E_{gap}/2e$, the
current crosses zero at some gate voltage (\mbox{$V_g \sim
-2.5$\,V} in Fig.~\ref{Fig3}). At this value of $V_g$ the electron
and hole packets contain the same number of particles, resulting
in zero net current. If $V_g$ is made more positive from this
point, the electron packets become bigger than the hole packets so
there is a net positive current. Above some $V_g$ the threshold
condition for hole injection is not met and the current is carried
entirely by electron packets. Likewise, as $V_g$ is made more
negative, the hole packets become bigger and the net current is
negative, and beyond some point it is carried only by holes.

According to the model, pumping should be absent in some interval
of $V_g$ when the SAW amplitude is below $\Phi_{SAW} <
E_{gap}/2e$. This is indeed the case, as can be seen in
Fig.~\ref{Fig4} \cite{difference}. By adjusting the position of
the bands with respect to the Fermi level (using $V_g$), one can
reach a threshold for the injection of either electrons or holes,
but simultaneous transport of both electron and hole packets
should not be possible at any $V_g$.

Exceeding the threshold is a necessary but not sufficient
condition for effective charge injection. A strong enough electric
field must also be present at the source contact to make the
tunneling barriers transparent for electrons and holes. As the
same condition determines the dc bias $V_{sd}$ required to drive a
dc current (Fig.~\ref{Fig2}) we can compare the SAW electric field
to that due to an applied bias voltage. In our device a dc current
flows above a bias of 1.5 V (Fig.~\ref{Fig2}a) giving an average
electric field of \mbox{$\sim 1.5$\,V/$\mu$m} along the nanotube.
The maximum electric field due to the SAW at \mbox{$P_{RF} = +20
$\,dBm} is indeed close to this value at \mbox{$E_{max} = 2 \pi
\Phi_{SAW}/ \lambda \sim 1.3$\,V/$\mu$m}. We note that the actual
electric field (caused either by dc bias or by the SAW) can be
significantly enhanced in a region very close to a thin contact
strip due to geometrical factors.

The maximum number, $n_{max}$, of electrons (holes) that can be
contained by a SAW potential well is given approximately by
$n_{max} = C\Phi_{SAW}/e$, where $C$ is the self capacitance of a
section of the nanotube containing the charge packet. Given a
ratio \mbox{$e/C \sim 5$\,mV}, expected for a carbon nanotube of 1
$\mu$m \cite{McEuen}, we arrive at $n_{max} \sim 40$,
corresponding to \mbox{$I_{SAW} = n_{max} ef \sim 20$\,nA} for
\mbox{$f = 3.2$\,GHz}. Thus, we can justify currents up to this
order as occurring by the transport mechanism described.

We have until now limited our discussion of the SAW-induced
current to the range of $V_g$ in which the dc conductance is zero.
Outside this range the current shows distinct peaks, after which
it drops dramatically, Fig.~\ref{Fig3}(c). We suggest that outside
the insulating region the free carriers screen the SAW field, the
carriers are delocalized, and the description of transport in
charge packets does not apply. In this `drag' regime the dc
current in the nanotube is proportional to the momentum transfer
from the wave to the nanotube's electronic system. If we assume,
by analogy with the situation in a 2DEG \cite{Weinreich,Wixforth},
that the momentum transfer decreases when the nanotube's
conductivity exceeds some value, then it explains the decrease of
the current with $V_g$ in Fig.~\ref{Fig3}. In this picture the
peaks in current in Fig.~\ref{Fig3}(c) manifest a crossover from
the transport in packets to the drag-type transport regime. This
assertion about the change of the transport mechanism can be
tested since the charge packet transport should be little affected
by a dc bias, whereas outside the insulating region when free
carriers exist in the nanotube, the effect of a bias is expected
to be much stronger. Such behavior is indeed observed as shown in
the inset of Fig.~\ref{Fig3}(c).

We conclude with a discussion on the feasibility of pumping single
electrons in carbon nanotubes. Whereas the Thouless mechanism is
more likely to be observed in metallic nanotubes, the classical
mechanism of SAW single electron pumping should be observable for
the semiconducting nanotubes studied here. In that mechanism the
number of electrons in a packet is fixed within an interval of
values of the relevant parameters, such as gate voltage or SAW
amplitude, due to Coulomb interactions. This work shows that
transport of charge in the form of packets can be realized in
nanotubes. For single-electron pumping each packet (or moving
quantum dot) should be populated with the same number of electrons
and the conditions at the source contact where the SAW potential
well is filled with electrons (holes) become crucial.

In particular, the time for electrons (holes) to enter the moving
SAW dot should be long compared to the $RC_{cd}$ time of the
junction, where $R$ is a contact resistance and $C_{cd}$ is the
capacitance between the contact and the dot. As an order of
magnitude estimate of the tunneling rate we take values \mbox{$R
\sim 10$\,M$\Omega$} and \mbox{$C_{cd} \sim 10$\,aF}, as deduced
from measurements of the device in the Coulomb blockade regime.
This yields an inverse $RC$ time of 10 GHz which only slightly
exceeds the SAW frequency of 3.2 GHz and implies that the number
of electrons tunneling into a SAW dot fluctuates from cycle to
cycle. The conditions at the source contact can be improved by
using a better (less resistive) source contact, such as Pd
\cite{Javey}. An alternative solution would be to induce electrons
or holes in a section of a nanotube adjacent to the source
electrode with the help of an additional side gate. Then, as with
the GaAs SAW pumps \cite{Talyanskii1}, the filling of a SAW
potential well with carriers will not involve a slow tunneling
process and the quantized transport regime should be observable.

The authors acknowledge stimulating discussions with L.S. Levitov,
D.S. Novikov, B.D. Simons B. Spivak, and D.J. Thouless. This work
was supported by the EPSRC UK (grant GR/R67521) and the EC program
SAWPHOTON. VT and PL acknowledge support from the Newton Trust and
the National Physics Laboratory, respectively.




\begin{thebibliography}{99}
\bibitem{diameter}
Given the relatively large diameter, it cannot be excluded that
the nanotube actually consists of a small bundle, or a multiwalled
nanotube. However, the characteristic rhomboid shown in Fig.
\ref{Fig2} suggests that we are dealing with a single nanotube.
\bibitem{devices}
Similar data to those reported here have been measured on two
other devices.
\bibitem{bandbending}
We estimate the band bending due to a mechanical perturbation of
the nanotube by the SAW to be much smaller than that due to the
SAW electrostatic field.
\bibitem{difference}
There is a slight difference between $I_{SAW}$($V_g$) graphs for
\mbox{$P = +20$\,dBm} in Figs. \ref{Fig3} and \ref{Fig4}. We
attribute this to a minor change in transducer resonant frequency
that resulted in a small decrease of SAW amplitude for the graph
in Fig. \ref{Fig4}. The presented interpretation of the data is
unaffected.
\bibitem{potential}
The source and drain potentials are effectively grounded at SAW
frequencies. This distorts the bands close to the contacts when
the SAW potential is non-zero, affecting panel $E$, but not panels
$D$ and $F$.
\end{thebibliography}
\end{document}